\documentclass{iau}
\usepackage{graphicx}
\usepackage{multirow}
\usepackage{natbib}
\usepackage{url}
\usepackage{hyperref}

\pubyear{2022}
\volume{369}  
\jname{The Dawn of Cosmology \& Multi-Messenger Studies with Fast Radio Bursts}
\editors{E.~Keane, K.~Rajwade, A.~Fialkov \& C.~Walker, eds.}

\title[The Petabyte Project] 
{The Petabyte Project}

\author[Lewis \textit{et al.}]   
{Evan F. Lewis$^{1,2}$, Sarah Burke-Spolaor$^{1,2}$, Maura McLaughlin$^{1,2}$, Duncan Lorimer$^{1,2}$, Kshitij Aggarwal$^{1,2}$, Devansh Agarwal$^{1,2}$, Joseph Kania$^{1,2}$, Nate Garver-Daniels$^{1,2}$, Joseph P. Glaser$^{1,2}$}

\affiliation{$^1$West Virginia University, Department of Physics and Astronomy, P. O. Box 6315, Morgantown, WV, USA  \\[\affilskip]
$^2$Center for Gravitational Waves and Cosmology, West Virginia University, Chestnut Ridge Research Building, Morgantown, WV, USA}

\begin{document}

\maketitle

\begin{abstract}
Transient radio sources, such as fast radio bursts, intermittent pulsars, and rotating radio transients, can offer a wealth of information regarding extreme emission physics as well as the intervening interstellar and/or intergalactic medium. Vital steps towards understanding these objects include characterizing their source populations and estimating their event rates across observing frequencies. However, previous efforts have been undertaken mostly by individual survey teams at disparate observing frequencies and telescopes, and with non-uniform algorithms for searching and characterization. The Petabyte Project (TPP) aims to address these issues by uniformly reprocessing data from several petabytes of radio transient surveys covering two decades of observing frequency (300 MHz--20 GHz). The TPP will provide robust event rate analyses, in-depth assessment of survey and pipeline completeness, as well as revealing discoveries from archival and ongoing radio surveys. We present an overview of TPP’s processing pipeline, scope, and our potential to make new discoveries.
\keywords{fast radio bursts, pulsars, radio transient sources, surveys, machine learning}
\end{abstract}

\firstsection 
\section{Introduction}

Sub-second-duration bursts of radio emission are known to come from a wide variety of astrophysical sources, including extragalactic Fast Radio Bursts (FRBs) \citealt{Lorimer07, Petroff22} and pulsars, some of which emit sporadically; the most sporadic pulsars are known as rotating radio transients (RRATs), which give isolated bursts every $N$ rotations, and intermittent pulsars, which become persistent emitters for limited amounts of time \citealt{RRATs, intermittent}. These radio transients are often discovered in large-scale radio surveys. Radio signals are also dispersed by the interstellar medium, creating a time delay between pulse arrival times measured across the observing bandwidth. It is therefore necessary to de-disperse the raw data at a large number of trial dispersion measures (DMs) to search for new transients; this is a computationally intensive process, and many legacy surveys were not de-dispersed to a sufficiently high maximum DM to detect high-DM FRBs. In addition to the unique telescope parameters and human-made radio frequency interference (RFI) environments at each observatory, each survey uses its own RFI mitigation and candidate classification algorithms. In most past surveys, there has been a manual inspection step that involves a human inspecting potentially hundreds of thousands of candidates by eye. These past searches have been effective at driving forward the study of FRBs, however have an unknown level of completeness, as this factor is not usually tracked; some search procedures have also been demonstrated to be insensitive to certain types of FRBs.

Many past searches do not account for these issues when reporting their event rates, introducing the possibility for missed bursts and inaccurate estimates of the population size and characteristics. It is especially important to accurately model the FRB event rate as a function of observing frequency, as an inaccurate estimate can lead to  difficulties in estimating the potential FRB detection rate of a new survey or instrument. 
We introduce The Petabyte Project (TPP), a large-scale uniform reprocessing of several petabytes of survey data across several telescopes and observing frequencies which will discover new transients and release the most sensitive FRB limits and frequency-dependent event rates to date.

\section{The Petabyte Project}
Table 1 summarizes our initial set of survey data that is currently available to TPP through public archival databases as well as memoranda-of-understandings with several large-scale radio survey groups. The data we have on hand amount to $\sim 2$ petabytes, covering three decades in frequency (300 MHz--20 GHz) and a wide variety of observatories. We will process all of this data uniformly with our pipeline (described in Section 3), which has been designed to read any dataset into a standard format and accurately detect transients, regardless of the observational or system parameters. 

\begin{table}
 \begin{center}
 \caption{TPP Surveys}\label{tpp-surveys}
 {\scriptsize\begin{tabular}{@{\extracolsep{\fill}}lcccccc}
 \hline
 Telescope (Survey) & Freq & BW & Sky Coverage & Max DM & Sensitivity & Reference \\
 & (GHz) & (GHz) & (hours deg$^2$) & (pc cm$^{-3}$) & (Jy) & \\
 \hline
 GBT (GBNCC) & 0.3 & 0.1 & 3880 & 3,000 & 0.63 & \citet{GBNCC} \\
 GBT (Drift-scans) & 0.3 & 0.1 & 135 & 1,015 & 0.783 & \citet{GBT350} \\
 Arecibo (AO327) & 0.3 & 0.06 & 114 & 1,095 & 0.783 & \citet{AODRIFT} \\
 GBT (GREENBURST) & 1.4 & 1 & 170 & 10,000 & 0.13 & \citet{GREENBURST} \\
 Arecibo (PALFA) & 1.4 & 0.3 & 56 & 9,867 & 0.06 & \citet{PALFA} \\
 Parkes (PMSURV) & 1.4 & 0.3 & 875 & 2,200 & 11.9 & \citet{PMSURV} \\
 Parkes (PHSURV) & 1.4 & 0.3 & 353 & 10,000 & 8.2 & \citet{PHSURV} \\
 Parkes (PASURV) & 1.4 & 0.3 & 350 & 1,500 & 7.3 & \citet{PASURV} \\
 Parkes (HTRU-mid) & 1.4 & 0.4 & 667 & 5,000 & 4.0 & \citet{HTRUmid} \\
 Parkes (HTRU-high) & 1.4 & 0.4 & 1625 & 5,000 & 4.6 & \citet{HTRUhigh} \\
 20-meter Telescope & 1.4 & 0.5 & 51 & 9,900 & 4.7 & \citet{20m} \\
 Parkes (P879) & 1.4 & 0.4 & 20 & -- & 0.46 & --\\
 GBT & 2 & 1 & 1.1 & -- & 0.047 & --\\
 GBT (Breakthrough) & 6 & 5 & 5.6 & 10,000 & 0.016 & \citet{BTL} \\
 Deep Space Network & 14 & 0.5 & $>$300 & -- & 0.03 & --\\
 Sardinia Radio Telescope & 6, 18.5 & 1 & $>$0.2,$>$0.01 & -- & 0.24, 0.6 & --\\
 \hline
     \end{tabular}}
    \end{center}
\vspace{1mm}
\scriptsize{
{\it Notes:}\\
The primary TPP data sets already in-hand, from the Green Bank Telescope (GBT), Arecibo Observatory (Arecibo), the Murriyang telescope operated by the Parkes Observatory (Parkes), the 20-meter telescope operated by the Green Bank Observatory, and others. Max DM refers to the maximum DM searched, if a single-pulse search was previously reported. }
\end{table}

Our uniform reprocessing will allow TPP to assess the sensitivity losses from each sub-project in our dataset and synthesize the results to provide a singular rate estimate for FRBs as a function of observing frequency. Because TPP's pipeline has been designed for homogenous processing, the FRB limits and event rates we release will be consistent in DM range limitations and will account for instrumental setups, and will allow for detection rate extrapolations for new surveys or instruments. 
Our processing and classification capabilities give TPP the capability to detect many transients which have been missed by previous efforts, including low-DM candidates flagged as RFI, high-DM sources in data which were not de-dispersed to a sufficiently high DM, and FRBs at $\geq$ 2 GHz, of which there have been few searches. We will also publicly release our large sample of new and re-detected candidates, creating a database of radio pulses from FRBs and radio pulsars across a range of observatories and frequencies. 

\section{Pipeline}
The wide variety of formats used to store observational data and metadata require TPP to have a robust method of handling any input data format, and a way to process it efficiently. We developed the python-based unified reader \texttt{your} \footnote{\url{https://github.com/thepetabyteproject/your}} to read a wide variety of single-dish-telescope data formats into a standard format with a uniform header. The unified reader allows for simultaneous read/write operations, meaning that we are able to search the data while writing it into the unified format, significantly reducing the computational load. An example candidate cutout generated by \texttt{your} is shown in Figure 1.

\begin{figure}
    \centering
    \includegraphics[scale=0.55]{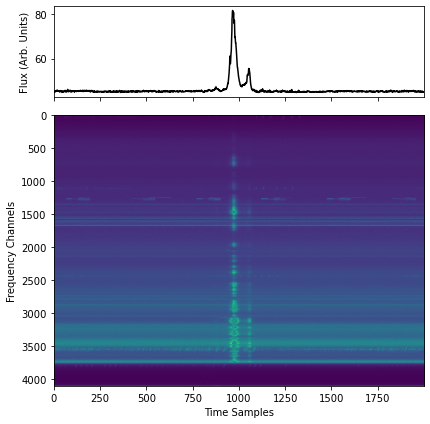}
    \caption{A candidate cutout of an FRB, generated by \texttt{your}. Top: the pulse profile averaged across the frequency band. Bottom: a frequency--time plot of the de-dispersed data. For each candidate we will upload candidate properties, visualizations, and any relevant metadata into our public database.}
    \label{fig:my_label}
\end{figure}

We will use an iterative injection-and-recovery search process to characterize our completeness for each survey, testing multiple RFI mitigation algorithms to choose the one that minimizes sensitivity losses for that survey. Once the completeness checks and RFI mitigation stages have been performed, we will search the data for transients using Heimdall\footnote{\url{https://sourceforge.net/projects/heimdall-astro/}}, a GPU-accelerated single-pulse-search program which been used to discover FRBs and single pulses from pulsars. To make sure that we perform a uniform-sensitivity search, all Heimdall search parameters will be fixed to conservative values for all the datasets: signal-to-noise ratio (S/N) threshold $\geq$ 6, de-dispersion up to 10,000 pc cm$^{-3}$, and pulse width $\leq$ 32 ms. The only exception will be for low-frequency surveys with short integration times, for which the maximum DM searched will be limited by the observation length.

Based on the candidate generation rate of our preliminary tests, we predict that TPP will generate $\sim$10 million candidates, producing a large bottleneck of labor and time required to assess each candidate. To classify all of our candidates, we developed \textsc{FETCH}\footnote{\url{https://github.com/devanshkv/fetch}}, a set of deep-learning based networks that produces a probability that the candidate is FRB-like, based on features identified in the intensity vs.\ frequency DM vs.\  time arrays. \textsc{FETCH} has been shown to have an accuracy and recall of $>99.5\%$, and has been used to accurately detect FRBs and other transients across the entire radio frequency band at observatories around the world. It has also been demonstrated to exceed the performance of traditional human-based identification methods \citep{121102}. 

At the conclusion of our processing and analysis, we will use \textsc{BARB}\footnote{\url{https://github.com/MorganWaddy/barb/}} to calculate the FRB event rate as a function of frequency. \textsc{BARB} uses the non-homogeneous Poisson process \citep{NPP}, which allows the detected instantaneous FRB rate to vary, allowing users to account for the variety of surveys and observations. \textsc{BARB} comes with the observing parameters and initially published detection rates for many of TPP's surveys, and can be augmented with any new surveys or discoveries, allowing TPP to expand our dataset and still easily incorporate the results to our final rate calculations.

\section{Potential Discoveries and Conclusion}
Our large-scale homogeneous reprocessing and accelerated de-dispersion gives us the potential to make several new discoveries in the archival data, including very high-DM FRBs which may be localized to high-redshift galaxies or be in dense environments. Our adaptive RFI mitigation algorithms will minimize the number of low-DM transients lost to aggressive RFI flagging, and allow us to potentially detect some of the nearest FRBs. Our use of \textsc{FETCH} will potentially unveil transients which were lost to insufficient matched filtering techniques in previous analyses, and allow us to estimate the fraction of transients potentially lost in the data. Furthermore, we will also correlate sub-threshold (S/N $< 10$) events at coincident positions, which may unveil new RRATs or repeating FRBs, potentially with detections across multiple observing frequencies and telescopes.

Any new transients discovered will contribute to the overall source population and be studied for potentially unique emission properties; TPP will also maintain a public database with frequency-time cutouts and relevant metadata for each astrophysical transient. In addition to new FRBs, we will collect a large sample of single pulses from intermittent pulsars and RRATs. These detections will allow us to discover new pulsars, perform a statistical analysis of the pulsar and RRAT populations, and systematically study the single-pulse properties of pulsars, such as nulling, mode changing, and subpulse drifting, across the majority of the radio observing band. As with all searches for transients, this reprocessing also contains the potential to unveil unique transients with previously unseen emission properties.

The discovery of new FRBs and our careful assessment of completeness will allow us to release a definitive census of FRB limits and rate measurements, transcending frequency, DM range, and instrumentation. Alongside these analyses, we will release a tool which can be used to predict the FRB detection rate for new instruments, allowing for the optimization of future FRB searches.

\acknowledgements
EFL and MAM are supported by NSF award AST-2009425.

\end{document}